\documentclass[aps, pre, twocolumn]{revtex4-1}
\usepackage{color}
\usepackage{amsmath,booktabs,array,amssymb,mathtools}
\usepackage{graphicx,threeparttable}
\usepackage{subeqnarray}
\usepackage{lineno}
\usepackage{cases}
\usepackage{tabularx}
\usepackage{tabu}
\usepackage{placeins}
\usepackage{hyperref}
\usepackage{fancyhdr}
\usepackage{endnotes}
\usepackage{amstext}
\usepackage{wasysym}
\usepackage{array,booktabs,tabularx}
\usepackage{multirow}

\begin{document}

\title{High speed imaging of solid needle and liquid micro-jet injections}
\author{Loreto Oyarte G\'alvez$^{1}$}
\email{l.a.oyartegalvez@utwente.nl}
\author{Maria Bri\'o P\'erez$^{1}$}
\author{David Fern\'andez Rivas$^{1}$}
\affiliation{$^{1}$Mesoscale Chemical Systems Group, MESA+ Institute and Faculty of Science and
Technology, University of Twente, P.O. Box 217, 7500 AE Enschede, Netherlands}

\begin{abstract} 
High speed imaging was used to capture the fast dynamics of two injection methods. The first one and perhaps the oldest known, is based on solid needles and used for dermal pigmentation, or tattooing. The second, is a novel needle-free micro-jet injector based on thermocavitation. We performed injections in agarose gel skin surrogates, and studied both methods using ink formulations with different fluidic properties to understand better the end-point injection. Both methods were used to inject water and a glycerin-water mixture. Commercial inks were used with the tattoo machine and compared with the other liquids injected. 
The agarose gel was kept stationary or in motion at a constant speed, along a plane perpendicular to the needle. The agarose deformation process due to the solid needle injection was also studied. The advantages and limitations of both methods are discussed, and we conclude that micro-jet injection has better performance than solid injection when comparing several quantities for three different liquids, such as the energy and volumetric delivery efficiencies per injection, depth and width of penetrations. A newly defined dimensionless quantity, the penetration strength, is used to indicate potential excessive damage to skin surrogates. Needle-free methods, such as the micro-jet injector here presented, could reduce the environmental impact of used needles, and benefit the health of millions of people that use needles on a daily basis for medical and cosmetic use.
%\newline
%\noindent \textbf{Keywords }
%\noindent Tattoo, jet injection, needle-free, ink
\end{abstract}

\maketitle

%%%%%%%%%%%%%%%%%%%%%%%%%%%%%%%%%%%%%%%%%%%%%%%%%%%%%%%%%%%%%%%%%%%%%%%%%%%%%
%%%%%%%%%%%%%%%%%%%%%%%%%%%%%%%%%%%%%%%%%%%%%%%%%%%%%%%%%%%%%%%%%%%%%%%%%%%%%
\section{Introduction} \label{s:Intro}

Tattooing, also known as dermal pigmentation, is done by inserting exogenous substances such as pigments into the dermis and leaving a permanent mark~\cite{sperry1991tattoos,kluger2012tattoos,islam2016medical}. The earliest evidence of tattooing procedures traces back to the fourth millennium BCE~\cite{friedman2018natural}. As evidenced by mummified skin, ancient art, and the archaeological records, tattoos have served two basic functions: medicinal or cosmetic \cite{tattohealth}. Two distinctive types of cosmetic tattoos exist; the conventional or purely decorative, and those intended to alleviate existing conditions and are defined as permanent make-up, e.g.\ scar camouflaging, alopecia or post-mastectomy pigmentation of a nipple on cancer patients. 

The societal acceptance of tattoos has varied over the years, however, there is a recent worldwide increase in its numbers and the social groups having tattoos or permanent make-up. According to a recent report, 12\% of the European population has one or more tattoos \cite{eurreport}. This corresponds to more than 44 million tattooed europeans, while the figures in USA and other countries is supposed to be similar or higher. This means that our society as a whole will be posed with scientific and technological challenges to reduce health risks caused by tattooing, and palliate its economic consequences.

Despite the advances made in electronics and improved hygienic conditions, the basic injection process has not changed much. According to experts, depending on the skin type and individual, a tattoo can be painful, cause skin related allergies, while 20-50\% of the ink is not injected~\cite{Petersen2016}. The method used for tattooing and permanent make-up is in principle the same: the repeated injection of one or several needles into the skin delivers ink droplets through the open wounds. The ink that adheres to the needle surface is transported into the dermis of the skin, as a function of the angle of the needle with respect to the skin, and the pressure applied by the tattoo artist or the cosmetic technician. Solid needles with single or multiple tips are typically sold as single-use consumables. The formulation of inks is kept as a secret by commercial brands, but their ingredients can be roughly categorised according to its function. Inks are composed of different pigments suspended in a carrier solution, together with binders and additives\cite{eurreport}. Each ink formulation has different ingredient proportions, conferring tailored fluid dynamic properties, such as viscosity, surface tension and density~\cite{jang2009influence,Petersen2016}. Pigments are inorganic particles responsible for the ink colour tone, with a particle size range of 0.1 $\mu$m - 50  $\mu$m. The larger the particle, the least they can be processed and removed by the immunological system of the human body, which in turn guarantees the permanence of the tattoo~\cite{Grant2015}. 

Less known is the fact that tattoo machines have been adapted for intradermal injections to evenly inject into a large area of the skin, dividing effectively the dose in smaller portions~\cite{Kim2017}.
During the last decades, alternative cutaneous delivery methods have been developed, including transdermal delivery injections at high pressures \cite{TABERNER} and needle-assisted jet injectors~\cite{LI2016195}. 
A liquid jet injector is a needle-free medical device that pressurises thin liquid filaments to penetrate the skin, with clear advantages over conventional injections with needles~\cite{Mitragotri:2006jd,Kim2017,Arora2017,munch2017dermal}. 
Jet injectors have helped in smallpox eradication, preventing rabies, influenza, malaria, hepatitis A and B, injecting insulin and analgesics, etc~\cite{BAXTER2005361,schramm2002transdermal,Gwen2017,Kim2017,Arora2017,munch2017dermal}. Some proven advantages are higher immunological response, drug dose sparing, reduction in pain with improved patient compliance, and reduction of accidental needle-stick incidents~\cite{hogan2015needle}.

To the best of our knowledge, a rigorous description of the tattooing process --one of the oldest transdermal injection methods-- is presented for the first time. A comparison with a novel micro-jet injector device --a needle-free injection method~\cite{carla}-- is performed on the basis of quantification of different observables. 

%%%%%%%%%%%%%%%%%%%%%%%%%%%%%%%%%%%%%%%%%%%%%%%%%%%%%%%%%%%%%%%%%%%%%%%%%%%%%
%%%%%%%%%%%%%%%%%%%%%%%%%%%%%%%%%%%%%%%%%%%%%%%%%%%%%%%%%%%%%%%%%%%%%%%%%%%%%
\section{Experimental setup and procedure} \label{s:MM}
%%%%%%%%%%%%%%%%%%%%%%%%%%%%%%%%%%%%%%%%%%%%%%%%%%%%%%%%%%%%%%%%%%%%%%%%%%%%%
\subsection{Solid needle injector} \label{ss:MMPmu}

\begin{figure}[h!]
	\includegraphics[scale=1]{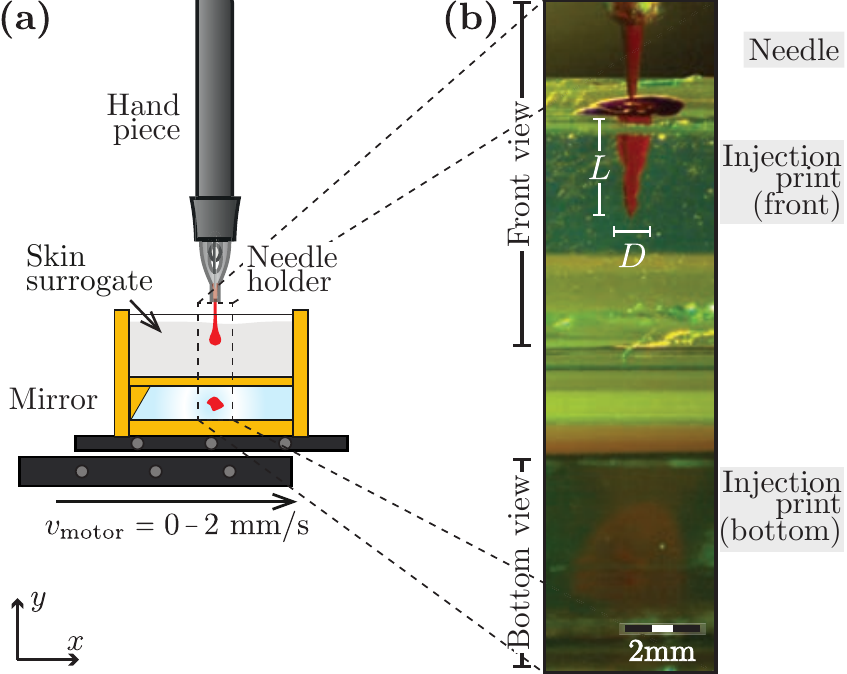}
	\centering
	\caption{(a) Schematics of the solid needle injector setup: the hand-piece of the tattoo machine is vertically fixed with a solid needle holder attached to it. The agarose gel skin surrogates were located below and almost touching the needle tip at rest. The agarose gel is kept stationary ($v_\text{motor}=0$) or in motion ($v_\text{motor}=2$~mm/s), along a plane perpendicular to the needle. A high-speed camera records the injection process at 1000~fps. (b) Example of the images obtained with the high-speed camera: the injection length $L$ and diameter $D$ are determined from the front view, whereas the bottom view shows the ink spread around the needle.}
	\label{fig:setup}
\end{figure}

A solid needle injector was vertically fixed in order to inject liquid into a skin surrogate made of agarose gel, as shown in figure~\ref{fig:setup}~(a). The solid needle injector is a conventional pigmentation instrument (PL-1000 Mobil, Permanent Line GmbH). It is composed of an electric control unit, a hand piece, and a consumable hygiene needle module. The hand piece works with an encased motor that moves the attached needle up and down in a smooth, cyclical pattern~\cite{machines}. The electric control unit enables an adjustable injection frequency, with nominal frequency values in the $f_n=[50-150]$~Hz range. The needle nozzle consists of a disposable single-use module attachable to the hand piece, which contains a sterilised solid needle, made from stainless-steel. Needles with a diameter of $D_\text{needle}$=4~mm were used for this study. 

The agarose gel samples were placed inside a custom-made holder, and the holder was located under the hand piece with the surface of the gel almost touching the needle tip, as shown in figure~\ref{fig:setup}~(a) and (b). The gel is confined by glass transparent walls providing a depth ($y$-axis), width ($x$-axis) and length ($z$-axis) of $\sim$3~mm, 3~mm and 20~mm, respectively. A mirror with a 45$^\circ$ orientation is placed below the gel. This system allows the simultaneous frontal and bottom visualisation of the injection processes into the gels. The holder is attached to a one-axis motorized translation stage (MT1/M-Z8, Thorlabs) which allows to keep the gel stationary or in motion at a constant speed $v_\text{motor}=2$~mm/s, along a plane perpendicular to the needle. 

Front view images were obtained using a color high-speed camera (Fastcam SA2, Photron) capturing 1000 frames per second; an example of the images acquired is shown in figure~\ref{fig:setup}~(b). In addition, the agarose gel deformation was measured with dry needles--without ink--, using a monochromatic high-speed camera (Fastcam SA-X2, Photron) recording at 10000 frames per second, and with a high-spatial resolution of 1409~pixels per millimeter.\\

%%%%%%%%%%%%%%%%%%%%%%%%%%%%%%%%%%%%%%%%%%%%%%%%%%%%%%%%%%%%%%%%%%%%%%%%%%%%%
\noindent\textbf{(i) \textit{Characterization of the needle displacement:}} \\
\indent In order to characterize the needle vertical displacement, we used the high-speed camera varying the nominal frequency $f_n$ from 50 to 150~Hz. The videos were recorded at 10000 frames per second, with a spatial resolution of 300 pixels per millimeter.  

\begin{figure}[h!]
	\includegraphics[scale=1]{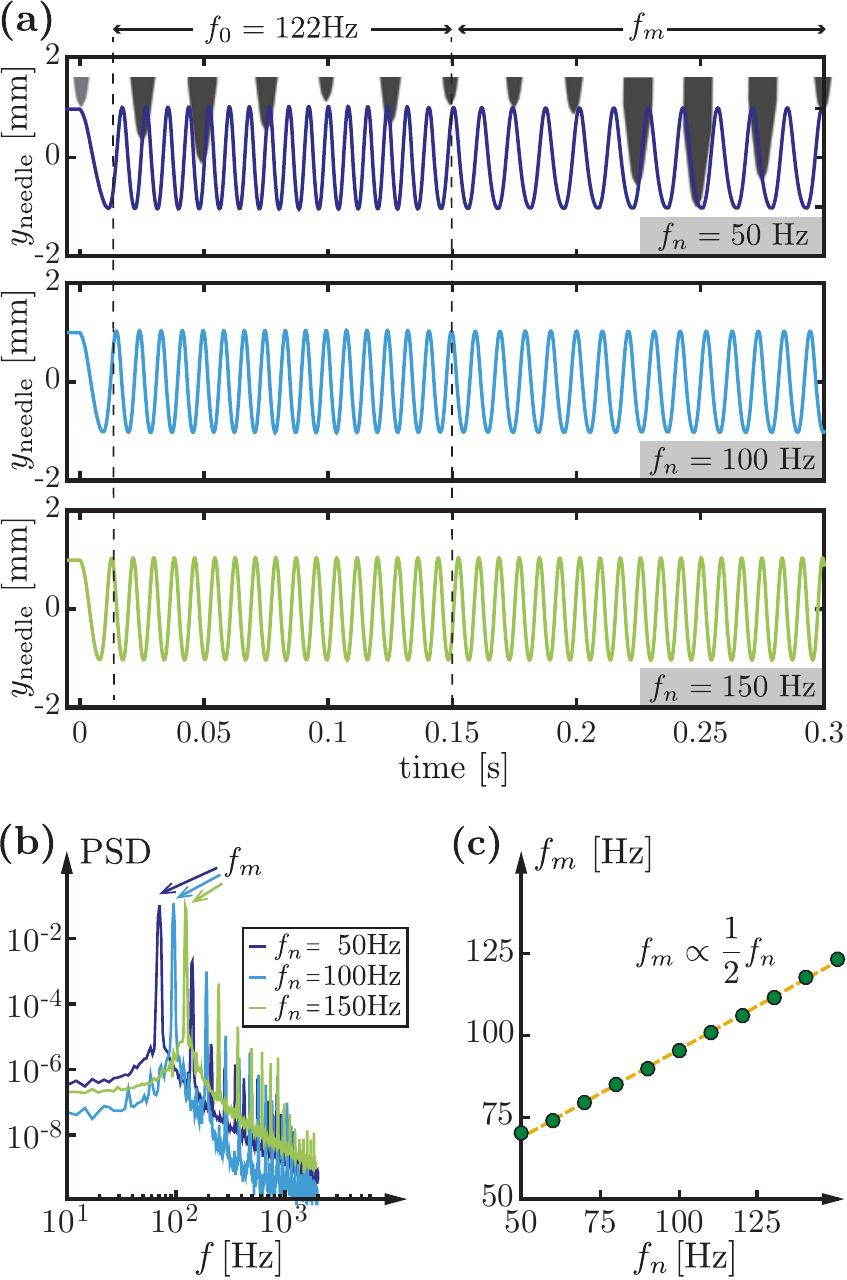}
	\centering
	\caption{(a) Vertical position of the needle tip versus time, for nominal frequencies $f_n=$50~Hz, 70~Hz and 90~Hz. During the first $\sim0.15$~s, the needle oscillations are the same in all cases. A slow oscillation is followed by a frequency $f_0\simeq$122~Hz, after which a stable measured frequency $f_m$ is reached in correspondence with the different nominal frequencies. (b) The power spectral density of the needle $y-$position for $t\gtrsim0.15$~s is plotted, for nominal frequencies $f_n=$50~Hz, 100~Hz and 150~Hz. The first peak corresponds to the measured frequency $f_m$. (c) The measured frequency $f_m$ versus the nominal frequency $f_n$ is also plotted. The fitted curve is represented by the dashed-line and shows a linear dependence $f_m\propto1/2f_m$.}
	\label{fig:fourier}
\end{figure}

The tip position is obtained and plotted with respect to time, as shown in figure~\ref{fig:fourier}~(a). During the first $\sim0.15$ seconds, the needle vertical displacement $y_\text{needle}$ is the same for every nominal frequency. A slow oscillation is followed by a cyclical displacement with constant frequency $f_0=122\pm1$~Hz and measured amplitude $a_m=1.018\pm0.006$~mm. After this time, $t\gtrsim0.15$~s, the measured frequency $f_m$ reach an stable value directly correlated to $f_n$. In figure~\ref{fig:fourier}~(b), we plot the power spectral density of $y_\text{needle}$ at $t>0.15$~s for the 3 cases shown in figure~\ref{fig:fourier}~(a), the position of the first peak corresponds to the measured frequency $f_m$. In figure~\ref{fig:fourier}~(c), we plot $f_m$ versus $f_n$ for all the experiments giving a linear relation $f_m=0.5f_n+42$.  \\ 

\begin{figure}[h!]
\includegraphics[scale=1]{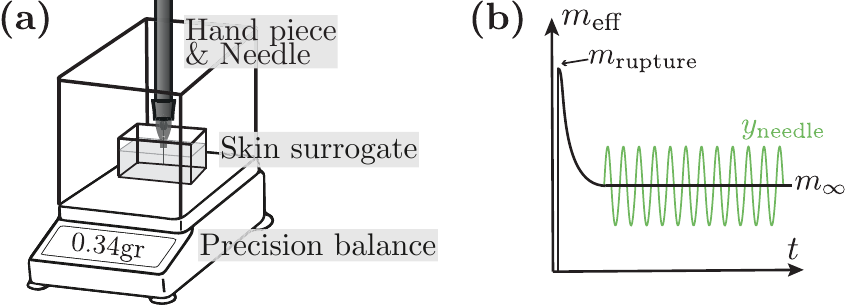}
\centering
\caption{(a) Schematic of the force measurement experimental setup. The skin surrogate is placed on the precision balance and the solid needle injector above it. (b) An initial peak mass $m_\text{rupture}$ is measured, corresponding to the rupture of the gel due to the needle penetration. After that, the balance measures an stable mass $m_\infty$, corresponding to the average mass applied by the oscillatory acceleration of the needle $\ddot{y}_\text{needle}$.}
\label{fig:Force}
\end{figure}

%%%%%%%%%%%%%%%%%%%%%%%%%%%%%%%%%%%%%%%%%%%%%%%%%%%%%%%%%%%%%%%%%%%%%%%%%%%%%
\noindent\textbf{(ii) \textit{Injection force measurement:}} \\
\indent In order to measure the force exerted by the solid needle injector into the agarose, we performed a separate experiment where the injector and skin surrogate setup were placed on a precision balance (Denver Instrument, APX-1000, $\Delta m$=0.1~mg), as shown in figure~\ref{fig:Force}~(a). The injector is switched on applying a normal force $F_\text{needle}$ on the agarose which is recorded by the balance, indicating the measured effective mass $m_\text{eff}$. The force and the effective mass are related by the equation
\begin{equation}
F_\text{needle}=m_\text{eff}|\vec{g}|,
\end{equation}
\noindent where $|\vec{g}|$ is the magnitude of the gravitational acceleration.

When the solid needle starts to move, an initial peak mass $m_\text{rupture}$ is measured. This mass corresponds to the moment when the agarose gel is ruptured, i.e. when the needle penetrates into the gel. After that, the measured mass reaches a stable value $m_\infty$, corresponding to the average effective mass due to the the oscillatory movement of the needle $y_\text{needle}$.

We have measured an effective mass $m_\text{rupture}=0.34\pm0.04$~gr and $m_\infty=0.16\pm0.02$~gr, for an agarose gel of 1\%wt and needle displacement frequency $f_m=70$~Hz ($f_n=50$~Hz), which corresponds to an applied normal force: $ F_\text{needle}^\text{rupture}=3.3\text{ mN}\,,\qquad F_\text{needle}^\infty=1.6\text{ mN}$. 
%\vspace{-0.2cm}
%\begin{equation*}
%F_\text{needle}^\text{rupture}=3.3\text{ mN}\,,\qquad F_\text{needle}^\infty=1.6\text{ mN}. 
%\end{equation*}

%%%%%%%%%%%%%%%%%%%%%%%%%%%%%%%%%%%%%%%%%%%%%%%%%%%%%%%%%%%%%%%%%%%%%%%%%%%%%
\subsection{Needle-free micro-jet injector} \label{ss:MMThC}
\begin{figure}[h!]
\includegraphics[scale=1]{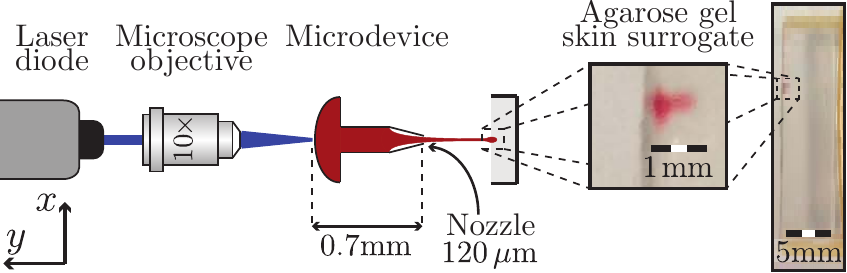}
\centering
\caption{Schematics of the needle-free micro-jet injector setup: A laser is focused at the bottom of a microfluidic device using a microscope objective. As a result, the bubble and jet are formed and are recorded using an ultrahigh-speed camera. The liquid jet penetrates the skin surrogate located in front of it. The image and the zoom-in insets show the agarose gel holder and the injection print, respectively, using a red colored glycerol-water mixture at $10\%$wt. }
\label{fig:setupJet}
\end{figure}

A continuous wave CW laser diode was focused at the bottom surface of a glass microfluidic device, which was partially filled with aqueous solutions of a molecular dye matching the laser wavelength. The liquid heats up above its boiling point in a few microseconds with an explosive phase transition resulting in a fast growing vapour bubble inside the microdevice; this phenomenon is known as thermocavitation~\cite{rastopov1991sound,Padilla-Martinez2014}. The bubble pushes the liquid forming a jet which in turn can penetrate into an agarose gel located in front, as shown in figure~\ref{fig:setupJet}. The microfluidic devices used were similar to those described elsewhere~\cite{berrospe2016continuous,carla}.

Microfluidic devices were designed and fabricated in glass substrates under cleanroom conditions. Each device has a fluidic chamber in which bubbles are created, and is connected to a tapered channel with a 120~$\mu$m diameter nozzle. For further details on this setup, the reader is referred to Berrospe \textit{et al}.~\cite{carla}
%The liquid is introduced through the chamber using capillary tubings connected to a precision glass syringe, and controlled by a syringe pump (Harvard PHD 22/2000).

The laser diode, with a wavelength $\lambda=450$~nm is focused at the bottom of the device with a 10$\times$ microscope objective. The spot has an elliptical shape, with beam diameters $r_x=33~\mu$m and $r_y=6~\mu$m and variable power P = 400-600~mW. 
The transparent glass walls  of a custom-made agarose holder permits the visualization of injection processes, as shown in the inset of figure \ref{fig:setupJet}. The agarose depth ($y$-axis), width ($x$-axis) and length ($z$-axis) are fixed at $\sim$5~mm, 3~mm and 24~mm, respectively.

The bubble growth, the liquid jet formation and the penetration into agarose slabs were recorded at $\sim400000$ frames per second using a ultrahigh-speed camera (Phantom v2640). The camera sensor is protected from the laser light using a colored glass filter centered at $\lambda=450$~nm.

%%%%%%%%%%%%%%%%%%%%%%%%%%%%%%%%%%%%%%%%%%%%%%%%%%%%%%%%%%%%%%%%%%%%%%%%%%%%% 
 \subsection{Liquid inks description}\label{ss:Injectate}
		  
Three different liquid inks were used on the experiments: commercially available Permanent make-up (PMU) inks, red dyed water and red dyed glycerin-water mixture at 10\%wt. By adding glycerin, the injection process, i.e. the adhesion to the needle tip and the diffusion in the agarose, is modified with respect to pure water. In order to maximize the absorbed energy by the liquid from the focused laser, in the needle-free micro-jet injector, the aqueous solutions are coloured using a red dye (Direct Red 81, CAS No. 2610-11-9) diluted at 0.5 \%wt.

The PMU inks are colloidal dispersions with flow-dependent viscoelastic properties. They are composed mainly by water and glycerol, and extra additives such as surfactants, solvents, binders and fillers~\cite{wijshoff2010dynamics,grande2016direct}. Pigments provide colour and due to insolubility in water, guarantee the permanent character of the injected ink. Additives are used in order to avoid pigment sedimentation while storage and help the re-dispersion of the fluid after it is used. Microbiological contamination is common in tattoos due to the high content of water an organic substances present on inks. In order to avoid the contamination, preservatives are added to the mixture. Other impurities that can be found on tattoos are primary aromatic amines (PAA) and  polycyclic aromatic hydrocarbons (PAH) \cite{eurreport,Petersen2016}. In this study, we use two organic PMU inks: PMU-black (Amiea, Organic line, Deep Black, MT. Derm) and PMU-red (Amiea, Organic line, Cranberry, MT. Derm).

\begin{figure}[h!]
\includegraphics[scale=1]{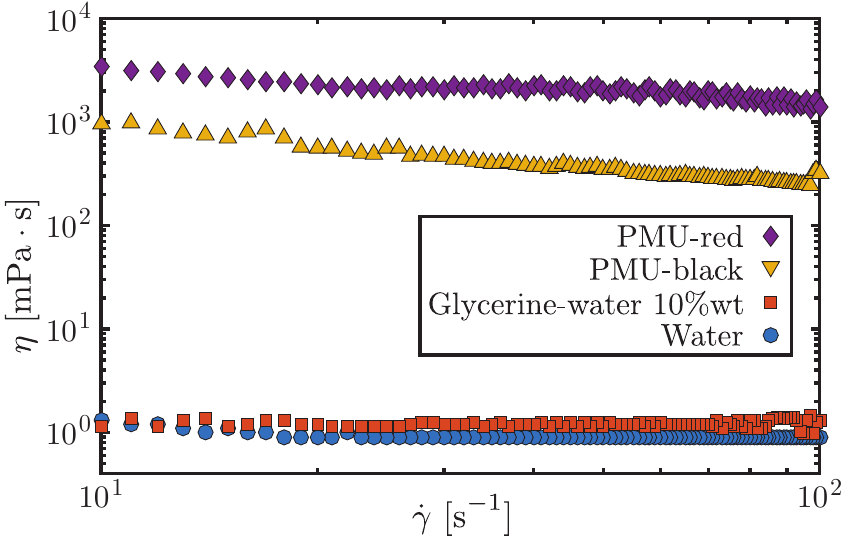}
  \centering
 \caption{The viscosity $\eta$ versus shear rate $\dot\gamma$ of the liquid inks: red dyed water (blue circles), red dyed glycerin-water 10\%wt (red squares), PMU-black ink (yellow triangles) and PMU-red ink (purple diamonds).}
 \label{fig:rheo}
\end{figure}

We performed rheology measurements of the liquid inks. Specifically, we have measured the viscosity $\eta$ varying the shear rate $\dot\gamma$ using a rheometer (Anton Paar MCR502) with a cone-plate geometry (with diameter $d=50$~mm and cone angle $\alpha=1^\circ$ ). As shown in figure~\ref{fig:rheo}, the red colored water and the glycerin-water mixture behave as Newtonian fluids with measured constant viscosity $\eta_\text{water}=0.9$~mPa$\cdot$s and $\eta_\text{glyc10\%}=1.2$~mPa$\cdot$s. On the contrary, the PMU inks show a shear thinning behaviour, i.e. the viscosity of the fluid decreases when the applied shear rate increases. The measured viscosity at high shear rates, $\dot\gamma\gtrsim100$~s$^{-1}$, for the PMU-black and PMU-red inks are $\eta_\text{PMU-black}\sim300$~mPa$\cdot$s and $\eta_\text{PMU-red}\sim2000$~mPa$\cdot$s, respectively.

The larger viscosity value of PMU inks is provided by the high concentration of pigment particles, and other additives not present in the colored water, and water-glycerin solutions. PMU inks were not used on the needle-free micro-jet injector due to visualisation limitations caused by the light absorption of the pigments that impeded the observation of thermocavitation.

%%%%%%%%%%%%%%%%%%%%%%%%%%%%%%%%%%%%%%%%%%%%%%%%%%%%%%%%%%%%%%%%%%%%%%%%%%%%%
\subsection{Agarose gel skin surrogate preparation} 

Among the skin surrogates widely used to simulate the mechanical properties of soft tissues, agarose is one of the most used due to its transparency which allows the optical quantification of injections~\cite{kendall2002delivery,deng2012preparation,schramm2004jet,williams2016jet,carla,Tagawa2}. The surrogates were prepared by diluting agarose powder (OmniPur agarose, CAS No. 9012-36-6.), in deionised water, with an agarose concentration of 1\%wt. The solution was heated up 45 seconds in microwave at full power. Once the phantoms were prepared, they were cooled down at room temperature for 5 minutes and stored at 4$^{\circ}$C.

%%%%%%%%%%%%%%%%%%%%%%%%%%%%%%%%%%%%%%%%%%%%%%%%%%%%%%%%%%%%%%%%%%%%%%%%%%%%%
%%%%%%%%%%%%%%%%%%%%%%%%%%%%%%%%%%%%%%%%%%%%%%%%%%%%%%%%%%%%%%%%%%%%%%%%%%%%%

\section{Solid needle injection method}\label{s:SolidNeedle}
In this section, we analyze the solid needle injector method in three stages: micro-indentation, stationary injection and moving injection. The micro-indentation process is one where the needle pushes the skin surrogate down without causing the rupture of the surrogate. The stationary injection is when the needle start the injection of the ink into the surrogate, immediately after the rupture occurs, keeping the gel fixed with respect to the injector hand piece. Finally, the moving injection process corresponds to a scenario closer to real life injection conditions, in which the needle is moved perpendicular to the skin surface. 

%%%%%%%%%%%%%%%%%%%%%%%%%%%%%%%%%%%%%%%%%%%%%%%%%%%%%%%%%%%%%%%%%%%%%%%%%%%%%
\subsection{Dynamic micro-indentation hardness test}\label{ss:indentation}
\begin{figure}[h!]
	\includegraphics[scale=1]{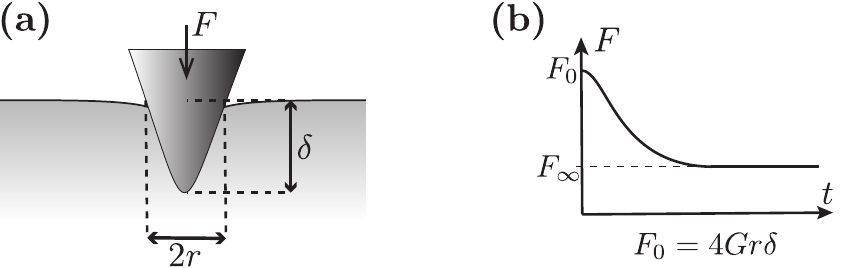}
	\centering
	\caption{(a) A schematic representation of the conical indentation test, where $\delta$ corresponds to the indenter tip position respect to the gel surface and $r$ is the radius of the impressed cone volume. (b) Force relaxation in time during the indentation test. An initial instantaneous force $F_0$ is exerted by the gel opposing the indenting force, which directly relates to the shear modulus $G$ as $F_0=4Gr\delta$.}
	\label{fig:ShearModulus}
\end{figure}

A micro-indentation hardness test is used to describe the hardness of a material to deformation with low applied loads. Recently, the interest to develop micro-indentation testing to characterize skin surrogate hydrogels has grown~\cite{Ahearne,YuhangHu,Ebenstein,CONSTANTINIDES,Hui,galli2009}. A specific indentation testing configuration is the conical indenter, in which the indenter is impressed into the surface of the agarose gel using a known applied force, as shown in figure~\ref{fig:ShearModulus}~(a). 

During indentation early stages (t $\le$ 1~ms), the water in the agarose gel does not have time to flow away and the gel behaves as an incompressible elastic solid. The instantaneous force $F_0$ correlates with the shear modulus $G$ of the gel~\cite{johnson_1985} as 
\begin{equation}
F_0=4Gr\delta, 
\end{equation}
\noindent where $\delta$ corresponds to the indenter tip position with respect to the gel surface and $r$ is the radius of the immersed indenter volume, as shown in figure~\ref{fig:ShearModulus}~(a). After this time, the solvent can flow away and the agarose gel starts to relax, the force reaches a threshold value $F_\infty$, and the gel behaves as a compressible elastic solid. The qualitative force behaviour is plotted in figure~\ref{fig:ShearModulus}~(b).

\begin{figure}[h!]
	\includegraphics[scale=1]{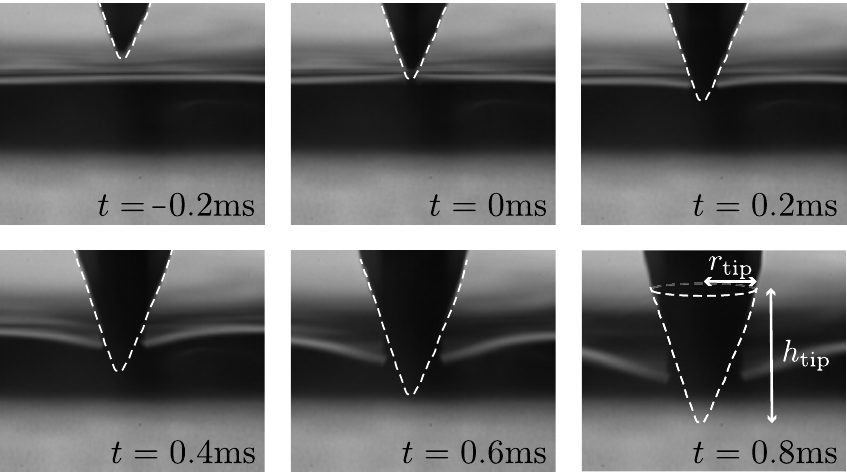}
	\centering
	\caption{Image sequence of the needle solid injection process before the skin surrogate rupture, the first needle-agarose contact occurs at $t$=0~s. The needle tip has a right circular conical surface, represented by the white-dashed line, with height and base radius $h_\text{tip}=0.174$~mm and $r_\text{tip}=0.068$~mm, respectively (Multimedia view).}
	\label{fig:IndentationSeq}
\end{figure}

In our experiment, the needle solid injector can be considered as a dynamic micro-indentation measuring instrument with a conical indenter. The solid needle travels rapidly and does not allow the relaxation of the gel, and as consequence, the measured force is the instantaneous force $F_0$ in relation to the needle displacement. The micro-indentation measurement will be valid before the rupture of the surface agarose gel occurs, typically before a millisecond. Furthermore, the tip of the solid needle has a right circular conical geometry, as shown in figure \ref{fig:IndentationSeq}, with height and base radius $h_\text{tip}=0.174$~mm and $r_\text{tip}=0.068$~mm, respectively. The needle tip position $\delta$, and the gel surface deformation are obtained image sequences of the experiment.

\begin{figure}[h!]
	\includegraphics[scale=1]{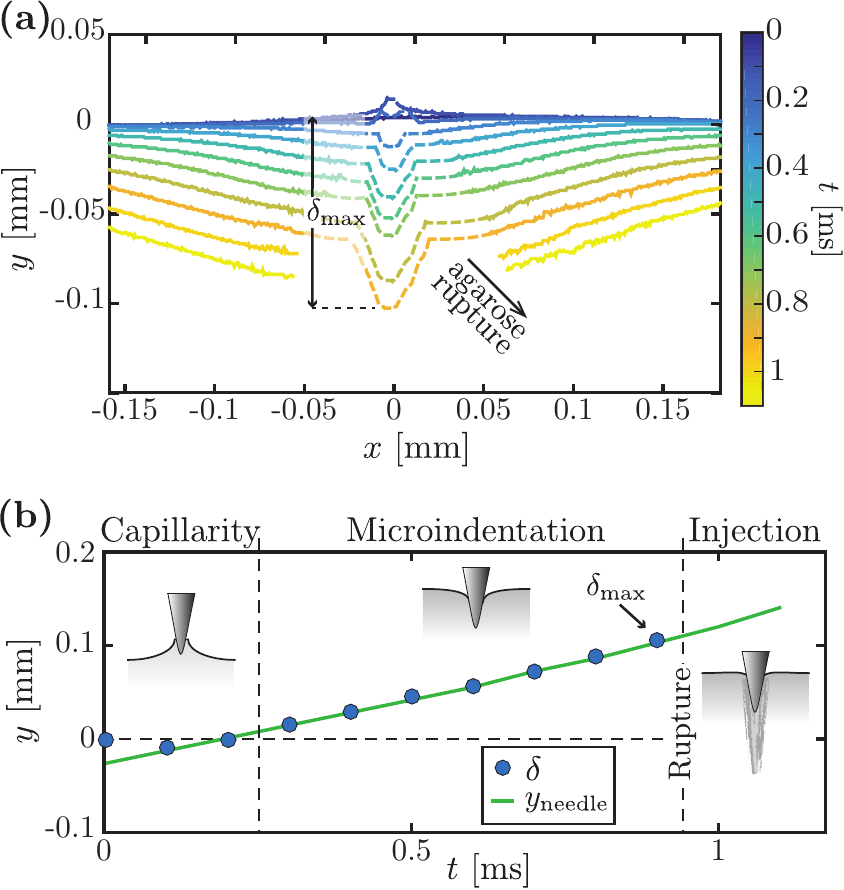}
	\centering
	\caption{(a) Skin surrogate surface deformation during solid needle micro-indentation. The maximum deformation just before rupture $\delta_\text{max}$  is shown. (b) Maximum deformation $\delta$ and expected needle tip position $y_\text{needle}$ versus time. The three stages of the surface deformation (capillarity, micro-indentation and rupture) are sketched and delimited.}
	\label{fig:Indentation}
\end{figure}

The gel deformation during indentation process is plotted in figure~\ref{fig:Indentation}~(a), where the zero position is defined in the initial agarose gel surface plane, i.e. the surface before injection. We observe three stages in the surface deformation: capillarity, indentation and injection. For a time $<2$~ms, when the needle is approaching, the first contact is with a water film in the surface of the gel, formed due to the environmental humidity and evaporation. At the contact point, a liquid bridge wetting the needle is created due to the capillary action, and the visualized deformation is negative $\delta<0$. The liquid bridge formation occurs too quickly to be captured in greater detail by our high-speed camera, however this phenomenon has been reported to play an important role in, for example, nano- and micro-indentation and AFM microscopy~\cite{CHEN2008,Men2009}.   

After that, when $y_\text{needle}\geq0$ and $t\geq0.2$~ms, the needle pushes the agarose gel down and the dynamic micro-indentation process starts. Finally, the force applied by the needle against the agarose is high enough to induce gel rupture, and the ink delivery, effectively starting the injection process. 

The maximum deformation $\delta$ caused by the needle tip displacement is plotted in figure~\ref{fig:Indentation}~(b), blue dots. The expected needle position without the agarose gel $y_\text{needle}$ is taken from the calibration process described in section~\ref{ss:MMPmu}, green line. As we expected, $\delta$ and $y_\text{needle}$ are in perfect agreement, which means the force exerted by the agarose gel is negligible compared to that exerted by the injector, meaning that the needle displacement is not affected by the gel. 

In the micro-indentation stage, the maximum deformation measured is $\delta_\text{max}=0.1062$~mm which corresponds to the applied force $F_\text{rupture}=3.3$~mN. We compare this force with $F_0$ as, \( F_\text{rupture}=3.3~\text{mN}=4G\delta_\text{max}r_\text{max}=F_0\), 
%\begin{equation}
%F_\text{rupture}=3.3~\text{mN}=4G\delta_\text{max}r_\text{max}=F_0,
%\end{equation}
\noindent where $r_\text{max}=r_\text{tip}\delta_\text{max}/h_\text{tip}=0.042$~mm. We can calculate the shear modulus $G$ of the skin surrogate, which represents the hardness, rigidity or stiffness of the material, obtaining \( G=185~\text{mN}/\text{mm}^2=185~\text{kPa}\).
%\begin{equation}
%G=185~\frac{\text{mN}}{\text{mm}^2}=185~\text{kPa}.
%\end{equation} 
This value matches the shear modulus range, $\sim$[30-3000]~kPa, reported in the literature for the agarose concentrations used~\cite{Normand,NAYAR}.

%%%%%%%%%%%%%%%%%%%%%%%%%%%%%%%%%%%%%%%%%%%%%%%%%%%%%%%%%%%%%%%%%%%%%%%%%%%%%
\subsection{Stationary injection}\label{ss:Stationary}

\begin{figure}[h!]
	\includegraphics[scale=1]{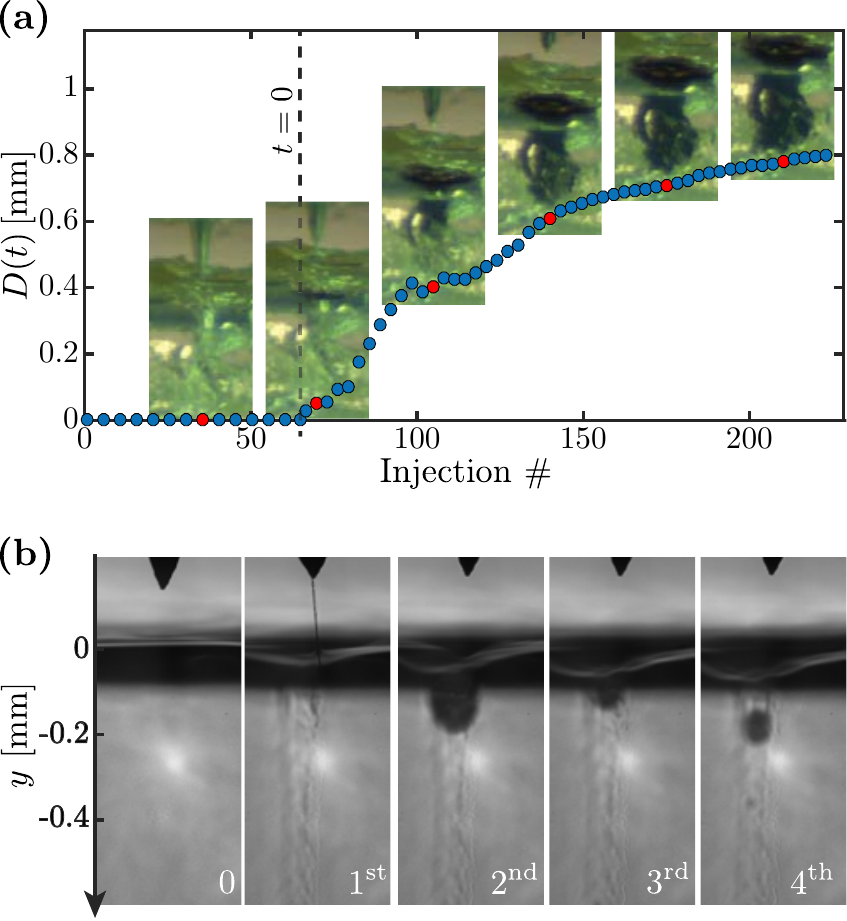}
	\centering
	\caption{(a) Injected-ink width $D(t)$ versus injection number. A clean needle holder is loaded with PMU-black ink delivering ink into the agarose after $\sim$60 injections; this moment is considered $t=0$ (dashed line). Around 100 injections later, the injection width reaches a threshold value $\sim0.8$~mm. Every image shows the injection at the time corresponding to the red point (Multimedia view). (b) High-resolution images show the skin surrogate agarose gel after 4 injections. The continuous deformation of the agarose gel surface is observed after each frame. A longitudinal hole of a diameter $\sim0.1$~mm remains in the agarose gel after every injection.}
	\label{fig:damage}
\end{figure}

Stationary injections were performed to understand the delivery process without the influence of specific factors, such as the needle injection angle, and translational speed with respect to the skin surrogate. The injection process in the case of a new needle holder loaded with PMU-black ink, is shown in figure~\ref{fig:damage}~(a). We observed that it takes over 50 injections for the ink adhered to the needle surface to slide down, and initiate the delivery into the agarose; this instant is considered $t=0$. After that, another 50 injections are needed to make a spot-width equivalent to the needle diameter $~0.4$~mm. Finally, around 100 injections later, the injection width reaches a threshold value of $\sim0.8$~mm. This plot shows that a unique injection is not enough to deliver a dose equivalent to the needle volume. Figure~\ref{fig:damage}~(b) shows a high-resolution image sequence of the agarose gel after four injections without ink. The deformation of the agarose gel surface after only four injections is clearly visible. Also the gel is internally damaged, and a longitudinal hole of a diameter $\sim0.1$~mm remains after one injection. After the second injection, a darker region inside the hole was observed, corresponding to a small water drop that came out of the agarose gel.

\begin{figure*}[t]
	\includegraphics[scale=1]{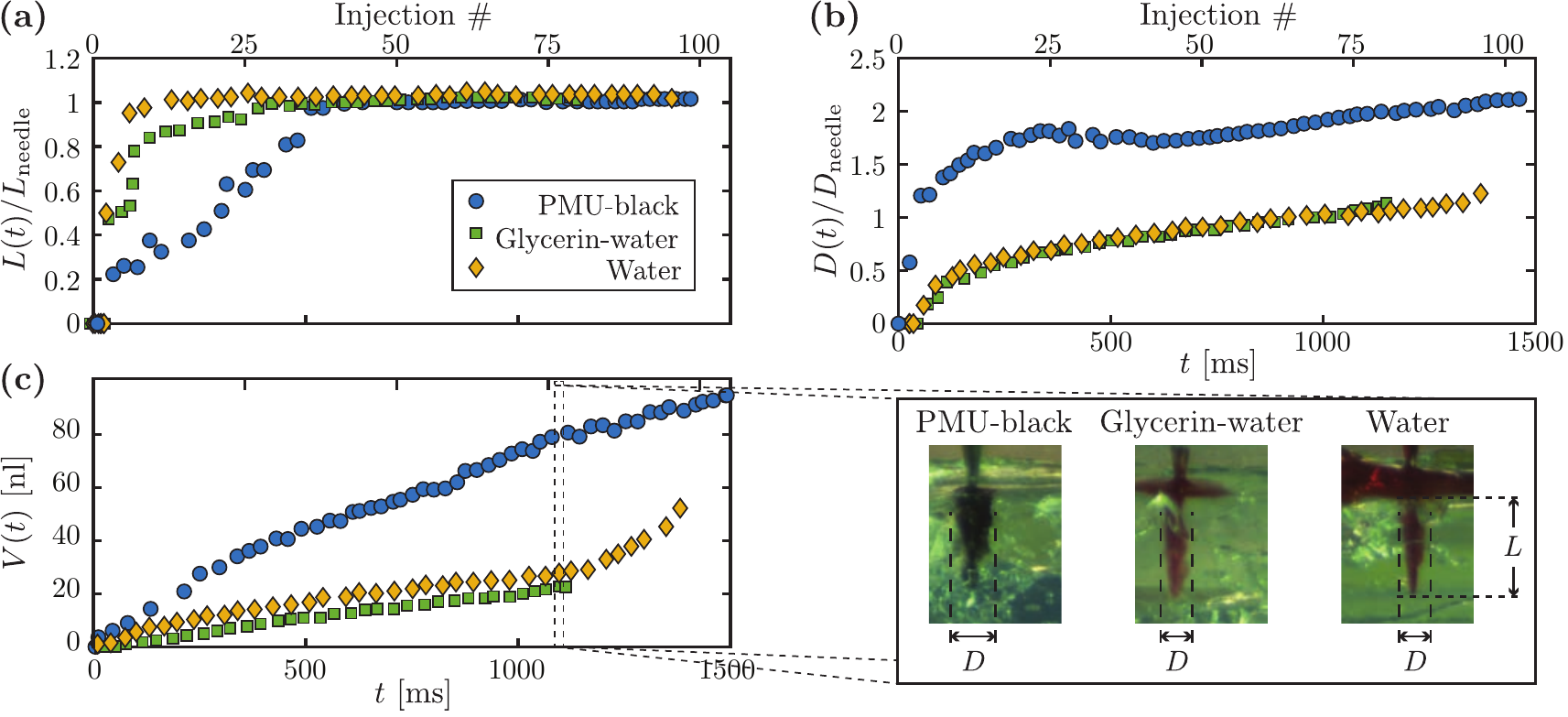}
	\centering
	\caption{(a) Injection depth and the immersed needle length ratio $L(t)/L_\text{needle}$, (b) Injection width and the inmersed needle diameter ratio $D(t)/D_\text{needle}$ and (c) Delivered volume $V(t)$, versus time for three of the inks: PMU-black ink, glycerin-water mixture and water. The image insets show the delivered dose at the same time $\sim1$~ms and/or the same injection number $\sim75$. The efficiency in volume delivery of PMU ink corresponds to the specific tailored properties for injection that the other formulations do not have. }
	\label{fig:StaticInjectionComparison}
\end{figure*}

The injection process, from t=0, is compared for three of the inks: PMU-black ink, glycerin-water mixture and water. We calculate the ratio between the injection depth and the immersed needle length $L(t)/L_\text{needle}$ and between the injection width and the immersed needle diameter $D(t)/D_\text{needle}$, as shown in figure~\ref{fig:StaticInjectionComparison}~(a) and (b). The total needle length is around 2~mm, but the total immersed needle into the agarose can vary in dependence of the initial distance between the needle tip and the agarose surface, mainly due to the agarose surface unevenness.  We observe that $L(t)/L_\text{needle}$ increases slower proportionally to the ink viscosity, but after $\sim40$ injections they all reach a threshold value $L/L_\text{needle}=1$. 

In the case of $D(t)/D_\text{needle}$, the spreading behaviour is the opposite, for the PMU ink the change from zero to one is almost instantaneous and  continues to increase up to $D/D_\text{needle}=2$ in around 100 injections. For the aqueous solutions, $D(t)/D_\text{Needle}$ saturates very quickly to 1, and the water and glycerin-water curves are almost indistinguishable (yellow diamonds and green squares in figure~\ref{fig:StaticInjectionComparison}~(b)). 

Assuming a circular conical injection shape, we estimate the injected volume as 
\begin{equation}
V(t)=\frac{1}{3}\pi \left (  \frac{D(t)}{2}\right )^2 L(t),
\end{equation}
\noindent and plotted in figure~\ref{fig:StaticInjectionComparison}~(c). The image insets show the delivered volume at the same time $\sim1$~ms or the same injection number $\sim75$. The plot shows that $V(t)$ for the PMU ink increases at least twice faster than for the aqueous solutions, reaching volumes of $\sim80$~nl and $\sim20$~nl respectively, as expected from the $L(t)/L_\text{needle}$ and $D(t)/D_\text{needle}$ plots. 

The large differences in behaviour are highly correlated with the differences in viscosity. For the solid needle used, with diameter $D_\text{needle}=0.4$~mm and an injection frequency and amplitude $f_m=74$~Hz, $a_m=1$~mm, the shear rate of the ink film wetting the needle is \vspace{-0.2cm}
\begin{equation}
\dot{\gamma}=\frac{v_\text{needle}}{D_\text{needle}}=\frac{a_m\cdot2\pi f_m}{D_\text{needle}}=1000\frac{1}{s}.
\end{equation}  

This means that the viscosities between the aqueous and the PMU inks are two order of magnitude different, from 1~mPa$\cdot$s to 300~mPa$\cdot$s. The PMU ink has a high adherence to the needle, making necessary to clean the needle in between experiments to prevent agglomeration. Moreover, the images show that the residual --not injected-- ink accumulates in the surface, making the solid needle injection method highly inefficient for low viscous inks. Furthermore, there is remarkable difference of the  liquid adhesion onto the needle when the viscosity  increases from $\eta_\text{water}=0.9$~mPa$\cdot$s to $\eta_\text{glyc10\%}=1.2$~mPa$\cdot$s. The images in figure~\ref{fig:StaticInjectionComparison}~(c) show this difference, where the amount of ink remaining in the agarose surface is considerably higher for pure water than glycerin-water mixture.

%%%%%%%%%%%%%%%%%%%%%%%%%%%%%%%%%%%%%%%%%%%%%%%%%%%%%%%%%%%%%%%%%%%%%%%%%%%%%
\subsection{Moving injection}
\begin{figure}[t]
	\includegraphics[scale=1]{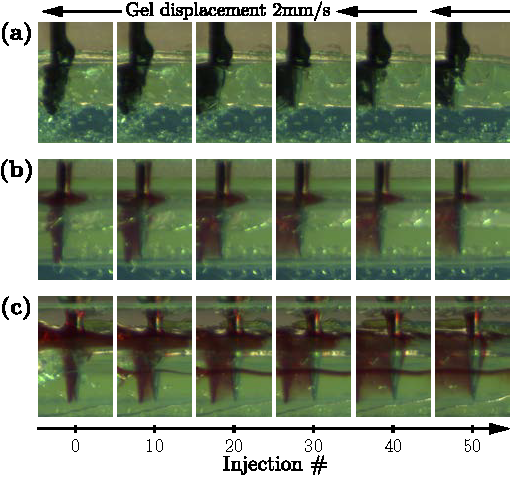}
	\centering
	\caption{Image sequences of the moving injection process for three inks: (a) PMU-black ink, (b) glycerin-water mixture and (c) water. The solid needle injector is initiated and the agarose gel is moving at 2~mm/s, from right to left. The spreading of the ink into the agarose is faster and wider for aqueous solutions than the PMU ink, which has a viscosity 200 times higher (Multimedia view).}
	\label{fig:dynamicpmu}
\end{figure}

A qualitative characterisation of conventional tattoo injection processes is performed by moving the translation stage at 2~mm/s orthogonal to the hand-piece. A total number of 50 injections, corresponding to $t\sim700$~ms, are studied. Figure~\ref{fig:dynamicpmu} shows a characteristic image sequence of the injection processes with different inks: (a) PMU-black ink, (b) glycerin-water mixture and (c) water. We observe that PMU ink has a much smaller spreading into the agarose gel, which we attribute to the several ingredients providing cohesion not present in aqueous solutions, i.e. glycerin-water mixture and water. In the case of the inks the surrogate saturation is faster and the ink oozes out on top of the agarose surface. This behaviour is expected because the agarose gel is mainly composed of water, therefore a maximum spreading is expected for the aqueous inks. Additionally, a residual volume of PMU ink remains adhered to the needle during the whole injection process. Pigment particles are the main responsible of the higher viscosity and density of PMU inks, giving such higher adhesion to the needle, and also limit the spreading into the agarose. In practice, the tailored composition and  resulting properties of PMU ink, facilitate the drying process, and the needle needs to be cleaned constantly to prevent agglomeration as described in section \ref{ss:Injectate}.

\begin{figure}[h!]
	\includegraphics[scale=1]{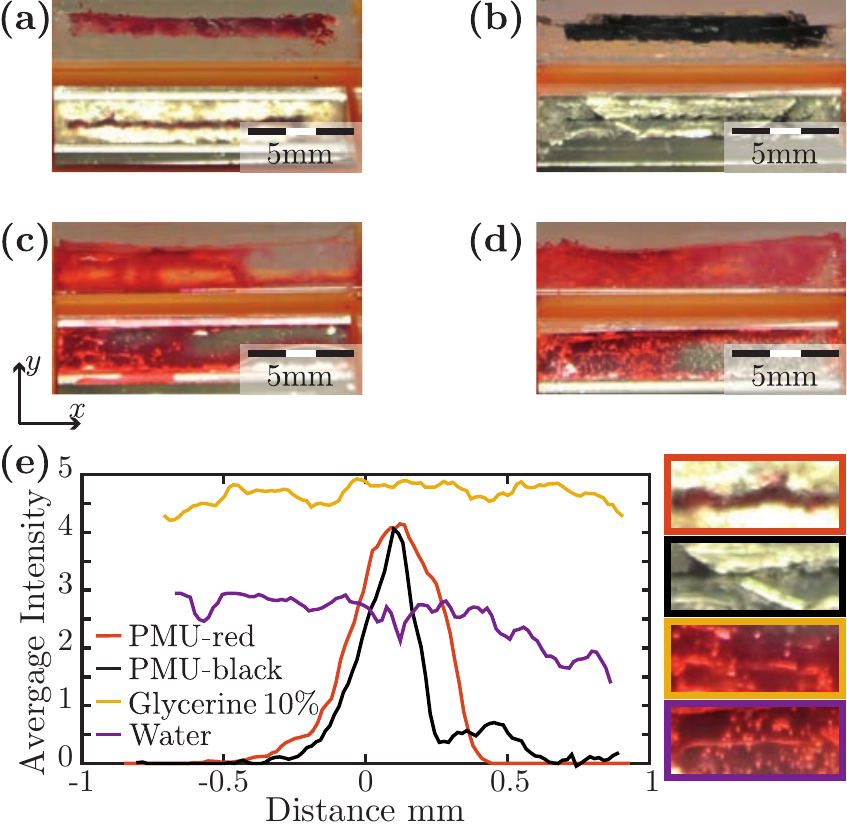}
	\centering
	\caption{Side and top view of post-injection skin surrogate for (a) PMU-red, (b) PMU-black, (c) glycerin-water mixture and (d) water inks. (e) Average intensity of the injected path observed in the mirrow. }
	\label{fig:pmuinjection}
\end{figure}

Images of the injected skin surrogate were taken immediately after the moving injection experiment. The agarose gel was cleaned before the imaging with a professional cleansing tonic (LaBina Aloe Vera tonic, PERMANENT-Line), in order to observe the post-injection result at the front and bottom, as shown in figure~\ref{fig:pmuinjection} for (a) PMU-red ink, (b) PMU-black ink, (c) glycerin-water mixture and (d) water. For the PMU inks, a well formed path line is visible in the front an bottom views. For the glycerin-dyed water solutions, the needle path is no longer visible due to its fast diffusion. However, we observed that the glycerin-water mixture presents less spread than pure water (see figure~\ref{fig:dynamicpmu}), in agreement with our observations described in section \ref{ss:Stationary}. The bottom view shows that this spread distributes asymmetrically around the needle, rather the liquid spreads depending on the micro-characteristics of the agarose gel, i.e. its porosity and inhomogeneities.  

Those observations are quantified in figure~\ref{fig:pmuinjection}~(e), the intensity of the injection path observed in the mirrow was calculated and averaged in the $x$ direction. A zoom-in of the mirrows for each ink is presented on the right. Two well defined peaks for the PMU inks are observed in the plot. The PMU-red peak is wider than the PMU-black, because the red ink is one order of magnitude more viscous, hence, the ink attaches better to the needle. The aqueous inks are completely spread in the agarose, showing an homogeneous average intensity. However, the glycerin-water mixture has a higher intensity, indicating that more ink has been delivered into the agarose.

%%%%%%%%%%%%%%%%%%%%%%%%%%%%%%%%%%%%%%%%%%%%%%%%%%%%%%%%%%%%%%%%%%%%%%%%%%%%%
\section{Needle-free micro-jet injection method}\label{s:NeedleFree}

The needle-free micro-jet injector exhibited different penetration characteristics from what we observed with the solid needle injector. The needle-free injector creates liquid micro-jets with a tip diameter $D_\text{jet}\sim50~\mu$m and a total ejected volume of ca.~50~nl, with a 100\% of the liquid delivered into the agarose for jet speeds larger than $\sim40$~m/s, which indicates no splash-back of liquid due to the sufficient kinetic energy of the jets. 

\begin{figure}[t]
	\includegraphics[scale=1]{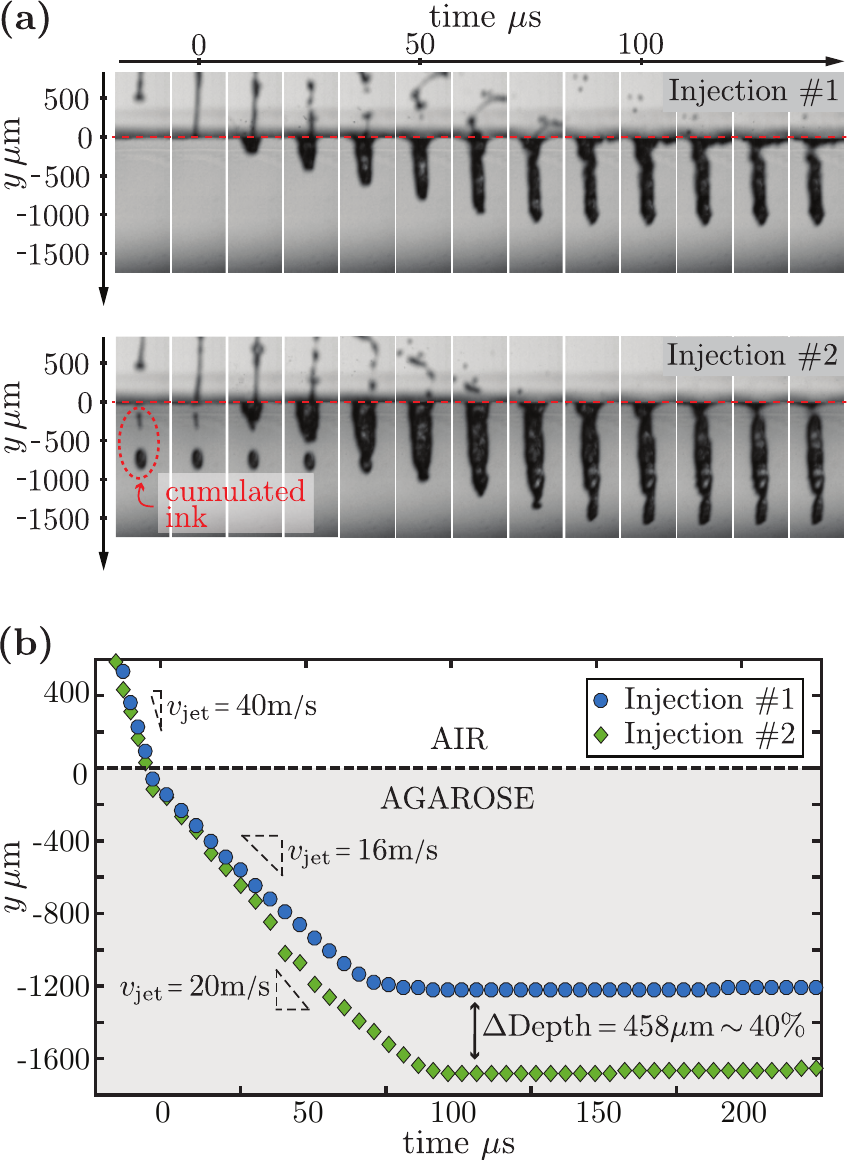}
	\centering
	\caption{(a) Two consecutive injections for pure water, with jet speed 40~m/s. The red-dashed line corresponds to the agarose gel surface. The cumulated ink from the previous injection is visible into the gel (Multimedia view). (b) Penetration depth versus time for the two continuous injections. Before the micro-jet reaches the agarose surface both jets have a speed of 40~m/s, with the impact, the first injection drops its speed to 16~m/s while the second injection to 20~m/s, penetrating around 40\% deeper into the agarose. }
	\label{fig:JetInjCum}
\end{figure}

Figure~\ref{fig:JetInjCum}~(a) shows image sequences of two successive needle-free injections of water jets entering the agarose at the same point. The penetration depths versus time for both injections are plotted in figure~\ref{fig:JetInjCum}~(b). Both micro-jet speeds were 40~m/s before impacting the agarose. In the first injection, the jet speed drops to 16~m/s, which means a decrease of 85\% in the micro-jet kinetic energy and is capable to penetrate 1~mm into the agarose. After the first jet opens a hole into the skin surrogate, each subsequent jet follows a micrometric longitudinal orifice into the gel. Therefore, the speed of the second micro-jet drops less than the previous one, up to 20~m/s, with its kinetic energy decreasing 75\% and allowing a penetration into the agarose 40\% deeper. 

\begin{figure}[t]
	\includegraphics[scale=1]{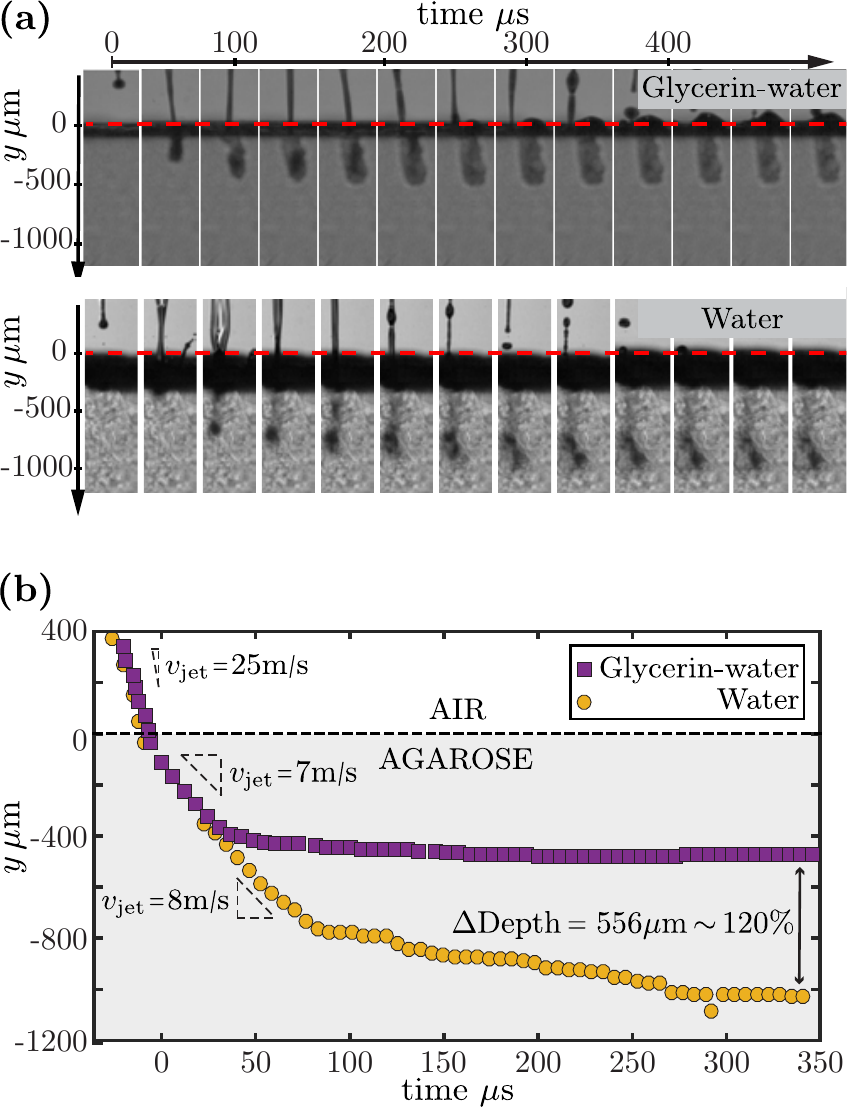}
	\centering
	\caption{(a) Image sequence of a single injection for pure water and glycerin-water mixture. In both cases, the micro-jet speed is $v_\text{jet}\sim25$m/s and the skin surrogate is agarose gel 1\%wt of concentration. (b) Penetration depth versus time for the two injections. The pure water ink is able to penetrate 120\% deeper than the glycerin-water mixture.}
	\label{fig:H2OvsGly}
\end{figure}

Experiments with glycerin-water mixture were performed in order to compare the injection process with the solid needle injection method. The shear rate experienced by the liquid jets is at least two order of magnitude higher than the shear rates provided by the solid needle injector. An image sequence of a single injection with micro-jet speed $v_\text{jet}\sim25$~m/s, $\dot{\gamma}=5\cdot10^5~1/$s, for pure water and glycerin-water mixture is shown in figure~\ref{fig:H2OvsGly}~(a). The differences observed in the agarose color and texture is an optical effect due to small illumination differences. It can be observed that, despite having the same initial jet velocities, water jets penetrate deeper than glycerin jets. The corresponding complete penetration events are plotted in figure~\ref{fig:H2OvsGly}~(b). 

We observed that inertial effects are more relevant in the early stages of jetting and penetration, while viscous dissipation gains prominence towards the end-point injection. The penetration speed of both jets right after entering the agarose drops to a comparable speed, 8~m/s and 7~m/s respectively. Afterwards, during jet deceleration, water jets penetrate much deeper than glycerin-water jets (120\% increase) due to the differences in the liquid viscosities of 25\% ($\eta_\text{water}=0.9$~mPa$\cdot$s and $\eta_\text{glyc10\%}=1.2$~mPa$\cdot$s). Though not visible in these figures, water jets spread (laterally) faster into the agarose gel, while the spread area of glycerin jets remains almost the same after each injection. 

%%%%%%%%%%%%%%%%%%%%%%%%%%%%%%%%%%%%%%%%%%%%%%%%%%%%%%%%%%%%%%%%%%%%%%%%%%%%%
%%%%%%%%%%%%%%%%%%%%%%%%%%%%%%%%%%%%%%%%%%%%%%%%%%%%%%%%%%%%%%%%%%%%%%%%%%%%%

\begin{table*}[t]
\caption{Comparison between solid needle and needle-free micro-jet methods. $S$ is the strength ratio between the injection pressure $p$ and the skin surrogate shear modulus $G=185$~kPa obtained in section~\ref{ss:indentation}. In the case of solid needle, the penetration depth $L$, width $D$ and volume $V$ are quantified for a single injection and for the end point.}
\begin{tabular}{|c|c|c|c|c|c|c|c|c|c|c|c|}
\cline{8-12} 
\multicolumn{1}{c}{} & \multicolumn{1}{c}{} & \multicolumn{1}{c}{} & \multicolumn{1}{c}{} & \multicolumn{1}{c}{} & \multicolumn{1}{c}{} &  & \multicolumn{3}{c|}{\#1 injection} & \multicolumn{2}{c|}{End point}\tabularnewline

\hline 
Method & Ink & $\, p$ [kPa]$\,$ & $\,S=p/G\,$ & $\, K$ {[}$\mu$J{]}$\,$ & $\,\epsilon_\text{energy}~\%\,$& $\, V_\text{0}$ {[}nl{]}$\,$ & $\, L$ {[}mm{]}$\,$ & $\,D$ {[}mm{]}$\,$ & $\,\epsilon_\text{vol}~\%\,$& $\,L$ {[}mm{]}$\,$ & $\,D$ {[}mm{]}$\,$\tabularnewline

\hline 
\hline
\multirow{3}{*} {\begin{tabular}{@{}c@{}}\textbf{Solid}\vspace{-0.15cm} \\ \textbf{ needle }\end{tabular}} & PMU-black & \multirow{3}{*}{2400} & \multirow{3}{*}{13} & 0.042 & 4.2$\cdot10^{-5}$ & 424 & 0.047 & 0.129 & 0.005 & \multirow{3}{*}{ $\sum\limits_{i=1}^{50}=1.5$} & 0.8\tabularnewline

\cline{2-2} \cline{5-10} \cline{12-12} 
 & Glycerin-water &  &  & 0.033 & 3.3$\cdot10^{-5}$ & 338 & 0.154 & 0.021 & 0.004 &   & 0.4\tabularnewline

\cline{2-2} \cline{5-10} \cline{12-12} 
 & Water &  &  & 0.027 & 2.7$\cdot10^{-5}$ & 283 & 0.197 & 0.019 & 0.004 &  & 0.4\tabularnewline

\hline 
\hline
\multirow{2}{*} {\begin{tabular}{@{}c@{}}\textbf{ Needle-free}\vspace{-0.15cm} \\ \textbf{ micro-jet }\end{tabular}} & Glycerin-water & \,3.2$\cdot10^{5}$\, & 1.73 & 2.85 & 5.7$\cdot10^{-2}$ & 8.9 & 0.450 & 0.192 & 88.1 & - & -\tabularnewline

\cline{2-12} 
 & Water & $3.1\cdot10^{5}$ & 4.32 & 16.16 & 3.2$\cdot10^{-1}$ & 20.2  & 1.2 & 0.205 & $\quad75.3\quad$ & $\sum\limits_{i=1}^2$=1.65 & 0.205\tabularnewline
\hline 
\end{tabular}
\label{t:Comparison}
\end{table*}

%%%%%%%%%%%%%%%%%%%%%%%%%%%%%%%%%%%%%%%%%%%%%%%%%%%%%%%%%%%%%%%%%%%%%%%%%%%%%
%%%%%%%%%%%%%%%%%%%%%%%%%%%%%%%%%%%%%%%%%%%%%%%%%%%%%%%%%%%%%%%%%%%%%%%%%%%%%

In contrast with the solid needle injector, no evident damage was observed on the agarose surface after multiple injections recorded with an image resolution of 150 pixels per millimeter. Moreover, the agarose shows a capacity of self-recovery as the path created by the micro-jet closes, and encapsulates the injected ink, see figure~\ref{fig:JetInjCum}~(a) first image, second injection.

%%%%%%%%%%%%%%%%%%%%%%%%%%%%%%%%%%%%%%%%%%%%%%%%%%%%%%%%%%%%%%%%%%%%%%%%%%%%%
%%%%%%%%%%%%%%%%%%%%%%%%%%%%%%%%%%%%%%%%%%%%%%%%%%%%%%%%%%%%%%%%%%%%%%%%%%%%%

\section{Comparison of injection methods}
The fact that no solid object is needed to rupture the surrogate with the needle-free micro-jet injector is a clear advantage in practical terms.  To make a fair comparison, the agarose gel surface rupture and damage, the spreading of the ink, among other observations are detailed in this section. The values calculated for all inks and both injection methods are presented in table~\ref{t:Comparison}, using the results in figure~\ref{fig:StaticInjectionComparison}, for the solid needle, and figure~\ref{fig:JetInjCum} and \ref{fig:H2OvsGly}, for the needle-free micro-jet. 

The electric energy supplied to the injectors is calculated as $E=P\cdot t$. The solid needle injector input power is $P=7$~W, and the time of a single injection is $t=1/70$~s, with a consumed energy $E=100$~mJ. The needle-free micro-jet injector input power corresponds to the laser power $P_\text{laser}=0.5$~W, and the corresponding time needed for the liquid to cavitate $t=1$~ms, corresponding to $E=5$~mJ. The kinetic energy transferred from the liquid to the skin is calculated as $K=1/2mv^2$, where the speed for the solid needle is $v_\text{SN}=a_m\cdot2\pi f_m$ and for the jet is $v_\text{jet}$. Therefore, the injection efficiency in terms of energy per injection is calculated as $\epsilon_\text{energy}=K/E \times 100~\%$. We have calculated that the needle-free micro-jet injector has a $\epsilon_\text{energy}$ three orders of magnitude higher than the solid injector.

The liquid volume deposited by the solid needle injector in each injection was estimated as the thickness of the liquid film around the needle tip:
\begin{equation}
\delta=c\left(\frac{\eta v}{\rho g}\right)^{1/2},
\end{equation}
\noindent where $\eta$ is the liquid viscosity, $\rho$ the liquid density, $v$ the flow velocity and $g$ the gravitational acceleration. This equation assumes that the flow velocity corresponds to the needle velocity, and that the surface tension role is negligible~\cite{LANDAU1988,RIO2017}.  The constant $c$ is taken as 0.8, which is a standard value for most Newtonian fluids. The ink volume around the needle is estimated as the volume of a hollow cylinder with inner radius as the needle radius $r_\text{inner}=D_\text{needle}/2$ and outer radius $r_\text{outer}=D_\text{needle}/2+\delta$.

The injection efficiency in terms of volume injected per injection event, we calculate it as the ratio between the deposited (solid needle) or ejected (needle-free micro-jet injector) volume $V_0$, and the volume remaining inside the agarose gel. For the former case, this is defined as the moment in which of solid needle retracts, and the latter is after the hole in the agarose closes. This efficiency is represented as  $\epsilon_\text{vol}=V_\text{inj}/V_\text{0} \times 100~\%$.  For the solid needle, the efficiency increases with each new injection because of the remaining ink from prior injections. There is, however, excess of liquid on the top of the agarose gel that does not penetrate at all, and contributes of the well known ink loss of 50\% in tattooing and permanent make up procedures (personal communication with several tattoing and PMU companies). The $\epsilon_\text{vol}$ is much higher for the needle-free micro-jet injector than the solid needle: five orders of magnitude.

In order to quantify the skin rupture and penetration characteristics, we compare the local stress induced by the solid needle and the jet impact, with a material-dependent critical local stress as proposed before~\cite{schramm2002transdermal}.  We define a new quantity, the penetration strength $S$, as the ratio between the injection pressure $p$ and the skin surrogate shear modulus $G=185$~kPa calculated in section~\ref{ss:indentation}. In the case of the solid needle, the injection pressure was calculated as the ration between the rupture force $F_\text{rupture}$ and the needle cross sectional area:
\begin{equation}
p^{\#1}_\text{needle}=\frac{F_\text{rupture}}{A_\text{needle}}=\frac{F_\text{rupture}}{\pi r^2_\text{max}},
\end{equation}
\noindent where $r_\text{max}=0.042$~mm is the radius of the inmersed needle volume at the moment of the maximum deformation of the skin surrogate, as explained in section~\ref{ss:indentation}. The pressure exerted by the jet onto the skin is $p_\text{jet}=\frac{1}{2}\rho v^2_\text{jet}$,
where $\rho$ is the density of the liquid, and $v_\text{jet}$ the jet speed~\cite{schramm2002transdermal}. 

The penetration depth $L$ and diameter $D$ in the latest stages or end-point of injection --after 50 injections for the solid needle, and after two injections for the needle-free injectors-- show two interesting results. First, that with only two jets is possible to reach the same depth as with solid needles. Second, that the injection resolution in spreading value ($D$), which corresponds to the practical level of detail in tattooing, is half for the jet injections. The current experimental setup for jet injections did not allow us to have the same reproducibility of conditions to reach larger number of injections. Contrary to the solid needle injections, each new jet carries a slightly higher velocity because of the reduction in liquid inside the microfluidic chamber. In future studies we will attempt to design a microfluidic device that ensures all jets in a sequence are ejected with the same velocity and total volume delivered.

%%%%%%%%%%%%%%%%%%%%%%%%%%%%%%%%%%%%%%%%%%%%%%%%%%%%%%%%%%%%%%%%%%%%%%%%%%%%%
%%%%%%%%%%%%%%%%%%%%%%%%%%%%%%%%%%%%%%%%%%%%%%%%%%%%%%%%%%%%%%%%%%%%%%%%%%%%%
\section{Conclusions} \label{s:conclusions}
We consider that this study will advance the knowledge of injection processes into soft substrates based on solid needles and needle-free micro-jet injectors. Our novel needle-free micro-jet injector employing thermocavitation is still in early development phases, however, the results indicate that the injection outcomes are superior to solid needle injectors in several aspects. Further work is required, particularly aimed at increasing the repeatability, frequency of injections, and overcoming challenges related to the use of commercial inks with unknown ingredients. Our results indicate that: 
\begin{itemize}

\item The power consumption of needle-free jet injections, 0.5~W, is an order of magnitude lower than the solid injector (7~W). This has practical relevance in future scenarios, where portability of injector devices is limited by the need of incorporating heavy batteries.

\item The calculated penetration strength S, reflects the proportional final damaged state of the skin surrogate. Values of S slightly above unity (jet injections) seem to correspond to an effective penetration with minimal damage, in contrast with irreversible skin surrogate deformations when S$\sim$10 (solid needle). 

\item The damage to the gel during the moving injections indicate that needle-free micro-jet injectors could have less negative effect than solid needles when injecting into skin. In real-life conditions, the displacement velocity and pressure applied with a hand-piece can fluctuate, damaging the skin because of the hardness needles.

\item The injection spreading increase with decreasing ink viscosity. However, the volume and energy efficiencies seem to increase proportionally to the viscosity of inks.

\item The injection efficiencies, $\epsilon_\text{energy}$ and $\epsilon_\text{vol}$, are higher for the needle-free micro-jet injector, with comparable end-point injections. The needle-free micro-jet injector in a single injection reaches penetration depths and widths  comparable to $\sim$50 injections with a solid needle. Even though multiple injections were not possible with the current needle-free micro-jet injector, the higher efficiency values demonstrate the superiority over the older solid needle method. For needle-free micro-jet injectors to compete with established injection methods in real-life scenarios, however, future designs should give multiple and reproducible injections, e.g.\ with multiple nozzles or  microfluidic geometries ensuring self-refill of ink after each jet. 
\end{itemize}

We conclude that needle-free micro-jet injections hold potential to mitigate health problems associated to needle-based injections, and avoid technical limitations of solid needle injections in medical treatments, e.g.\ vaccines, scar camouflaging, alopecia, etc. The reduction of environmental contamination with needle-free injections could be huge, while benefiting millions of people using needles on a daily basis for medical and cosmetic uses. With a long list of challenges ahead, we believe there will be a time, not far from now, when needle-free micro-jet injectors will be reliable, safer, and efficient liquid delivery platforms, and where needles will not be that much needed.

%%%%%%%%%%%%%%%%%%%%%%%%%%%%%%%%%%%%%%%%%%%%%%%%%%%%%%%%%%%%%%%%%%%%%%%%%%%%%
%%%%%%%%%%%%%%%%%%%%%%%%%%%%%%%%%%%%%%%%%%%%%%%%%%%%%%%%%%%%%%%%%%%%%%%%%%%%%
	
\section*{Acknowledgements}
We would like to thank Stefan Schlautmann and Frans Segerink for their technical support during fabrication and optical setup construction.
To Edgerton's centre for the use of the Phantom high-speed camera and illumination. The material support of PERMANENT-Line GmbH \& Co. KG. and MT-Derm is kindly acknowledged, as well as the practical and theoretical training offered by PERMANENT-Line GmbH \& Co. KG. 
DFR acknowledges the recognition from the Royal Dutch Society of Sciences (KHMW) that granted the Pieter Langerhuizen Lambertuszoon Fonds, 2016.

\section*{References}
\bibliography{bib}

\end{document}